\documentclass[apj]{emulateapj}

\def\xmm{{XMM-{\it Newton}}}
\def\chandra{{\it Chandra}}

\newcommand{\cgs}{ ${\rm erg~cm}^{-2}~{\rm s}^{-1}$} 
\newcommand{\lum}{\rm erg~s$^{-1}$}
\def\gtrsim{\mathrel{\hbox{\rlap{\hbox{\lower4pt\hbox{$\sim$}}}\hbox{$>$}}}}

\def\lum{{\rm {erg~s$^{-1}$}}}
\usepackage{graphics}
\usepackage{subfigure}
\usepackage{epsfig}
\usepackage{alltt}
\usepackage{verbatim}
\slugcomment{Accepted on April 10, 2012}

\shorttitle{\chandra\ HRC data of CID-42}
\shortauthors{F. Civano et al.}

\begin{document}

 \title{Chandra High resolution Observations of CID-42, a candidate recoiling SMBH}

\author{ F. Civano \altaffilmark{1}, 
M. Elvis \altaffilmark{1}, G. Lanzuisi\altaffilmark{1,2,3}, T. Aldcroft\altaffilmark{1}, M. Trichas\altaffilmark{1}, A. Bongiorno\altaffilmark{3},
M. Brusa\altaffilmark{3}, L. Blecha\altaffilmark{4}, A. Comastri\altaffilmark{5}, 
A. Loeb\altaffilmark{4}, M. Salvato\altaffilmark{6}, A. Fruscione\altaffilmark{1}, 
A. Koekemoer\altaffilmark{7}, S. Komossa\altaffilmark{6,8,9}, 
R. Gilli\altaffilmark{5}, V. Mainieri\altaffilmark{10}, E.
Piconcelli\altaffilmark{2,11}, C. Vignali\altaffilmark{12}}

   
\altaffiltext{1}{Smithsonian Astrophysical Observatory, 60 Garden St., Cambridge, MA 02138, USA}
\altaffiltext{2}{INAF-Osservatorio Astronomico di Roma, via Frascati 33, Monteporzio-Catone 00040, Italy}
\altaffiltext{3}{Max-Planck-Institut f\"ur extraterrestrische Physik, Giessenbach strasse 1, 85748 Garching, Germany}
\altaffiltext{4}{Department of Astronomy, Harvard University, 60 Garden Street, Cambridge, MA 02138, USA}
\altaffiltext{5}{INAF-Osservatorio Astronomico di Bologna, via Ranzani 1, Bologna 40127, Italy}
\altaffiltext{6}{Max-Planck-Institute for Plasma Physics,Excellence Cluster, Boltzmannstrass 2, Garching 85748, Germany} 
\altaffiltext{7}{Space Telescope Science Institute, 3700 San Martin Drive, Baltimore, MD 21218, USA}                                     
\altaffiltext{8}{Technische Universitaet Muenchen, James-Franck-Strasse 1/I, Garching 85748, Germany}
\altaffiltext{9}{Excellence Cluster Universe, Boltzmannstrass 2, Garching 85748, Germany}
\altaffiltext{10}{ESO, Karl-Schwarzschild-Strasse 2, 85748 Garching, Germany} 
\altaffiltext{11}{XMM-Newton Science Operations Centre, ESA, P.O. Box 78, 28691 Villanueva de la Cañada, Spain}
\altaffiltext{12}{Dipartimento di Astronomia, Universit\'a di Bologna, via Ranzani 1, Bologna 40127, Italy}


\begin{abstract}
We present \chandra\ High Resolution Camera observations of CID-42, a candidate recoiling supermassive black hole (SMBH) at z=0.359 
in the COSMOS survey. 
CID-42 shows two optical compact sources resolved in the HST/ACS image embedded in the same galaxy structure and a velocity offset 
of $\sim$1300~km/s between the H$\beta$ broad and narrow emission line, as presented by Civano et al. (2010). 
Two scenarios have been proposed to explain the properties of CID-42: a GW recoiling SMBH and a double Type 1/ Type 2 AGN system, where one of the two is 
recoling because of slingshot effect. In both scenario, one of the optical nuclei hosts an unobscured AGN, while the other one, either an obscured 
AGN or a star forming compact region. The X-ray \chandra\ data allow to unambiguously resolve 
the X-ray emission, and unveil the nature, of the two optical sources in CID-42. 
We find that only one of the optical nuclei is responsible for the whole X-ray unobscured emission observed and a 3$\sigma$ upper limit on the flux of 
the second optical nucleus is measured. The upper limit on the X-ray luminosity plus the analysis of the multiwavelength spectral energy distribution 
indicate the presence of a starforming region in the second source rather than an obscured SMBH, thus favoring the GW recoil scenario. 
However the presence of a very obscured SMBH cannot be fully ruled-out. A new X-ray feature, in a SW direction with 
respect to the main source, is discovered and discussed. 

\end{abstract}


\section{Introduction}

Coalescing SMBH pairs should give rise to very strong gravitational wave (GW) events in the
universe. After the merging event, the newly formed BH can experience a
recoil with respect to the center of the
galaxy due to asymmetric emission of gravitational radiation (Peres 1962, Bekenstein 1973).

Recent numerical relativity simulations have shown that the recoil velocity can be as high as 
a 5000 km/s for rare configurations of BH spins and mass ratio 
(Campanelli et al. 2007; Baker et al. 2008; Lousto et al. 2011a, 2011b). These recoil velocities are higher 
than the escape velocity of the host galaxy, so the SMBH will be ejected from the system and it may be detected as displaced quasar 
(Madau et al. 2004; Loeb 2007). 
Depending on the amount of material it can carry along, the ejected SMBH
will shine for 10$^6$-10$^7$ yr (Blecha et al. 2011), until it exhausts the fuel and is not recognizable as a displaced quasar anymore.
The recoil may significantly influence the SMBH and galaxy co-evolution (Guedes et al. 2011; Sijacki et al. 2011) and have 
implications for unified models (Komossa \& Merritt 2008).

In some cases, gas dynamics and stellar interactions may be insufficient to drive the SMBH binary to the GW-dominated regime 
of inspiral and merger, in which case the binary evolution may stall.
The arrival of a third galaxy, with its own SMBH, can induce hardening of the original binary by ejecting the newly arrived SMBH. 
This mechanism, usually referred as gravitational slingshot (Saslaw et al. 1974), produces 
one or more SMBHs displaced from the center of the same galaxy (Hoffman \& Loeb 2007). 

While extensive simulations have been performed (Gualandris \& Merritt 2008; Dotti et al. 2010; Blecha et al. 2011), 
observational searches for recoiling SMBHs are scarce (Bonning et al. 2007,
Komossa 2012 for a review).
 Two candidates have been found by spectroscopic searches for highly shifted
broad lines (Komossa et al. 2008, Shields et al. 2009). 
Another candidate has been discovered using spectropolarimetric observations (Robinson et al. 2010) and 
a fourth based on optical imaging with HST (Batcheldor et al. 2010). 
Based on X-ray imaging with \xmm, Jonker et al. (2010) found an unusual
X-ray source offset from the center of a nearby galaxy and offered 3 possible
explanations, one of which is a recoiling SMBH. 

 We (Civano et al. 2010; C10) studied the properties of CID-42 (z=0.359)
which is the only candidate to date to have both optical imaging and spectroscopic 
signatures of a recoiling SMBH. In this paper we present the results obtained using the full sub-arcsecond imaging power 
of the \chandra\ High Resolution Camera (HRC-I, 0.137$^{\prime\prime}$/pixel) to
unambiguously resolve the X-ray emission in CID-42. 
We assume H$_0$ = 70~km~s$^{-1}$~Mpc$^{-1}$, $\Omega_M$ = 0.3 and $\Omega_{\Lambda}$= 0.7.

\section{The properties of CID-42}

Out of 1761 X-ray detections in the \chandra\ COSMOS survey (Elvis et al. 2009; Civano et al. 2012 submitted), CID-42 is the only source 
that clearly shows two optical sources (SE and NW) separated by 0.5$^{\prime\prime}$ (2.5 kpc) in the 
HST/ACS image and embedded in the same galaxy. 
In C10 we presented detailed modeling with Galfit (Peng et al. 2002) of the optical surface brightness
of CID-42 based on high ($\sim$0.025$^{\prime\prime}/pixel$ resolution COSMOS HST/ACS (F814W) image
(Koekemoer et al. 2007) to decompose the emission in CID-42. The results showed that the SE optical source 
has a point-like morphology, typical of a bright Active Galactic Nucleus (AGN), while the NW 
optical source has a more extended profile with a scale length of $\sim$0.5 kpc. 
In the optical spectra of CID-42 (VLT, Magellan), a velocity offset of $\sim$1300 km/s is 
measured between the broad and narrow component of the H$\beta$ line (Figure 5 and 6 of C10). 

C10, linking the offset broad emission line and the point morphology of the SE
source, proposed two possible scenarios to explain the properties of CID-42: a GW recoiling SMBH, leaving no BH in the galaxy center; 
a Type 1/ Type 2 AGN system in the same galaxy where the Type~1 is recoiling due to the slingshot effect in a triple system. 
Given the point like morphology and the broad emission lines in the optical
spectra, the SE source hosts an active unobscured SMBH in both C10 scenarios.
The nature of the NW source, instead, is still not clear. 
In the GW recoil scenario, the SE SMBH is recoiling from the  
NW source, which represents the galaxy core from which the BH has been ejected. 
In the slingshot scenario, the NW source host an active SMBH, which may be an obscured AGN. 

An alternative scenario, reported by Comerford et al. (2009; Co09), proposes the
presence of a dual system in CID-42, where the two merging sources are Type 2 
AGN (from optical emission line diagnostics) spectroscopically resolved in CID-42, making use of a Keck/DEIMOS spectrum.
However, the presence of broad Balmer lines (H$\beta$ and H$\alpha$; see
the SDSS DR7 quasar catalog in Shen et al. 2011) in the
optical spectra of CID-42, does not support the Type 2 (with narrow optical
emission lines only) nature proposed by Co09 for at least one of the two
sources. 

 X-ray emission, being insensitive to obscuration up to N$_H\sim$10$^{24}$cm$^{-2}$, is a strong discriminant and can probe the 
active nature of the optical sources in CID-42 and reveal the actual physical
process taking place. 
Two X-ray emitting AGNs are expected in the slingshot scenario in C10 and the Co09 interpretation, while the SE source 
would be the only powerful X-ray emitter in the GW recoil scenario.

CID-42 was observed in the C-COSMOS survey with \chandra\ ACIS for $\sim$166 ks, but its position was at a large off-axis angle, where 
the degraded PSF did not allow the association of the X-ray emission with one or both the optical sources. 

\section{Chandra Observation}
\subsection{Data Reduction}
\begin{figure}
\fbox{{\includegraphics[width=0.5\textwidth]{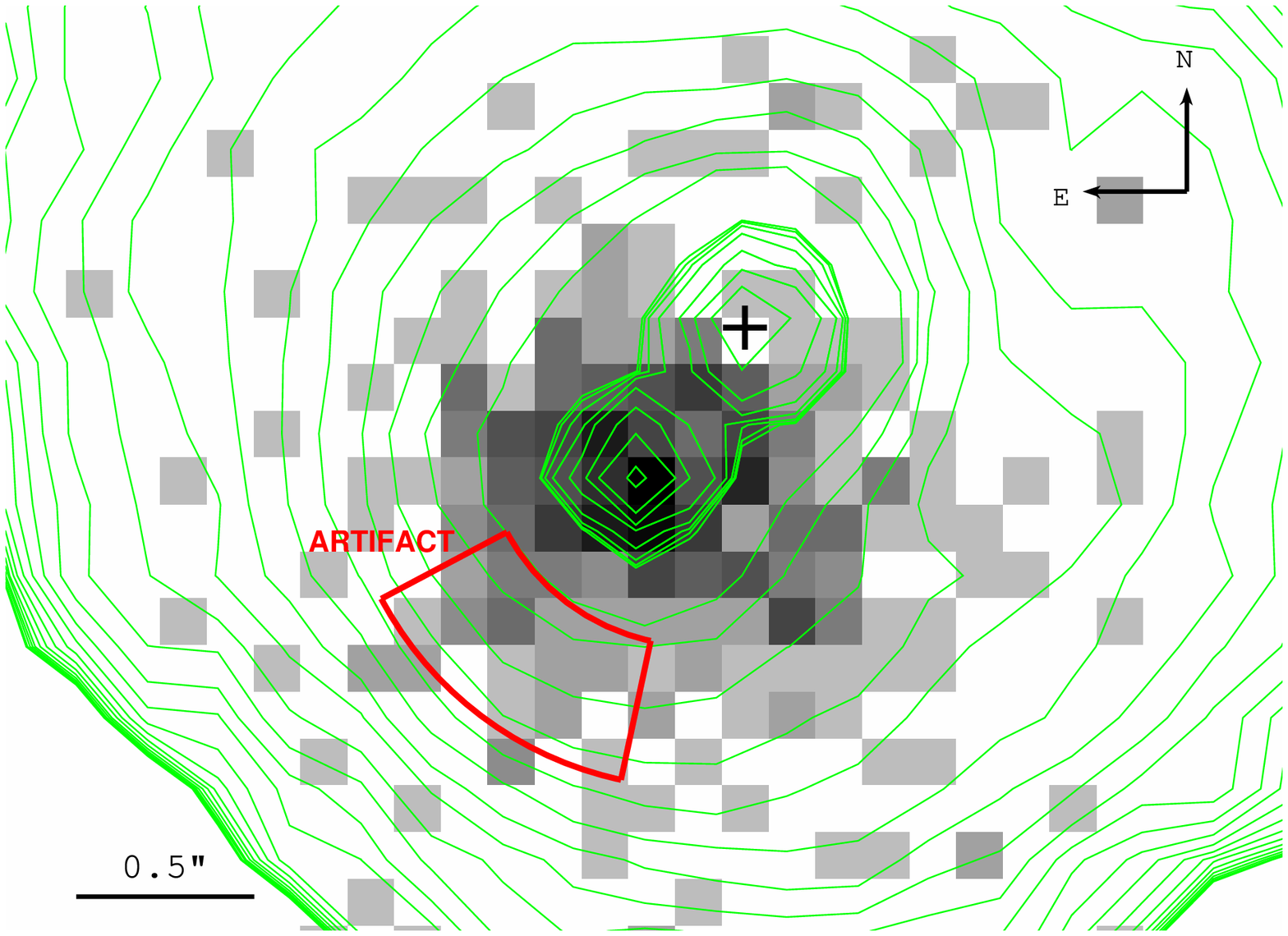}}}
\fbox{{\includegraphics[width=0.5\textwidth]{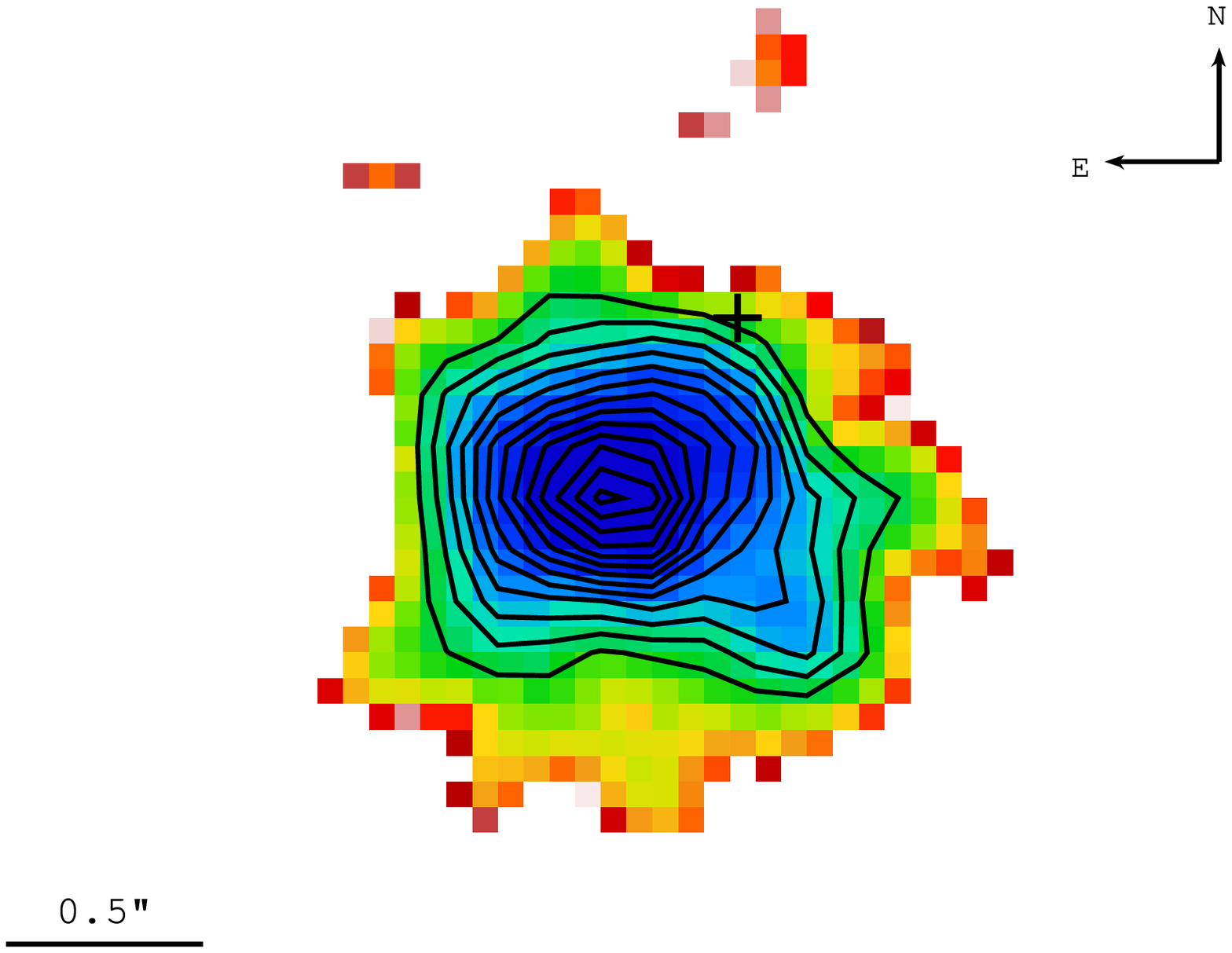}}}
\fbox{{\includegraphics[width=0.5\textwidth]{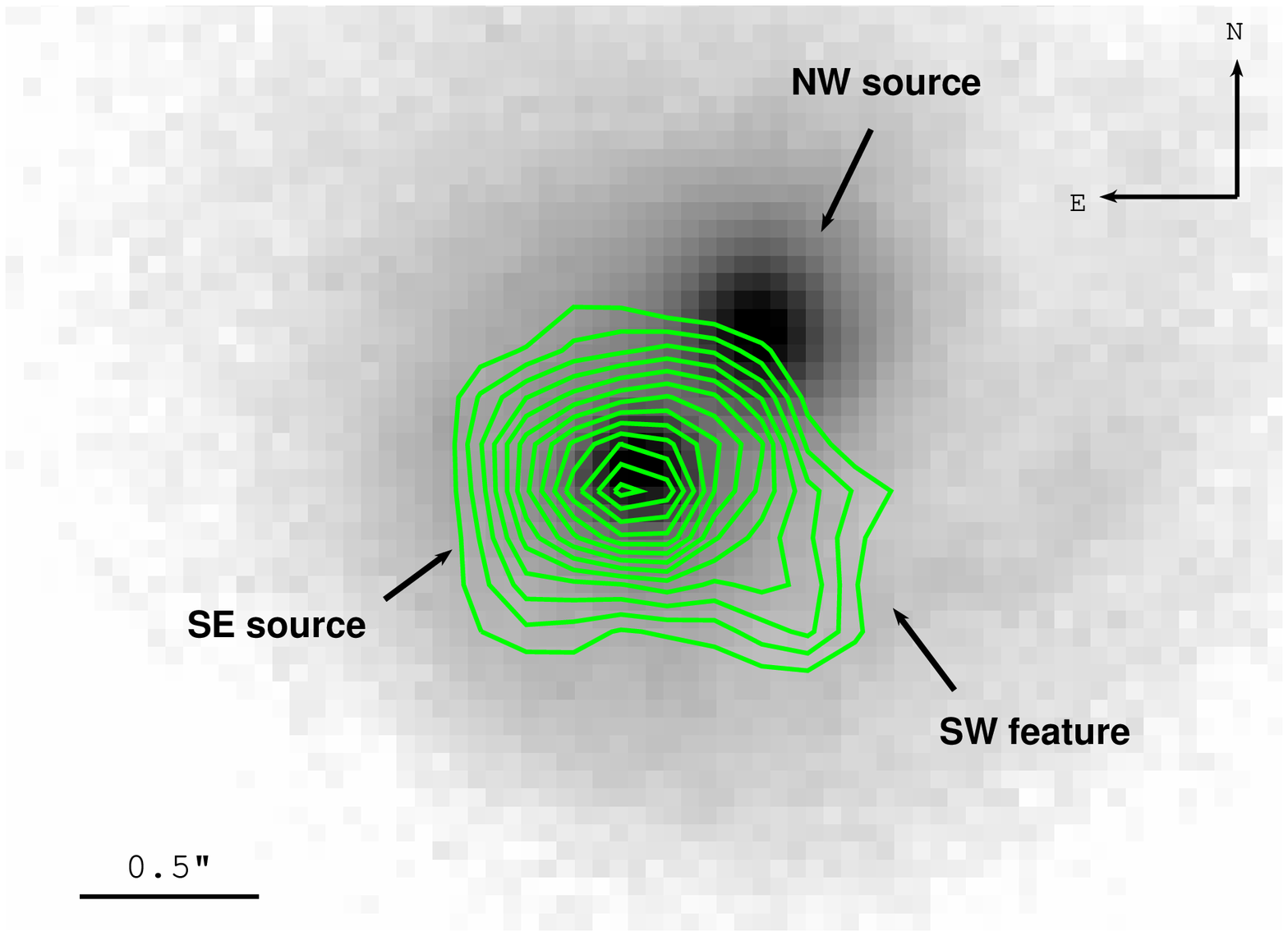}}}
 \caption{\small {\it Top:} \chandra\ HRC-I X-ray full band image with HST/ACS contours overlaid (green). 
The red solid line represents the HRC artifact region. The cross indicates the position of the NW source determined using the HST/ACS contours. 
{\it Middle:} Adaptively smoothed \chandra\ HRC-I X-ray full--band image with a 3 pixel radius Gaussian
kernel. X-ray contours are overlaid (black solid line). {\it Bottom:} 
HST/ACS (filter F814W) image with the X-ray (green) contours overlaid. The NW and SE optical 
sources have been labelled together with the SW feature. The three images are on the same WCS scale and relative astrometry correction has been applied. }
\label{image}
\end{figure}

CID-42 was observed on 2011 January 26 (during Cycle 12, observation ID 12810) with the \chandra\ 
HRC-I (Murray et al. 1997) camera for 79,3 ks.
The source was observed on-axis, just 15.2$^{\prime\prime}$ away from the aimpoint. 
The data were reprocessed using the \chandra\ 
software (CIAO 4.3) tool {\it chandra\_repro} and the latest calibration
files (CALDB 4.4.1). The light curve was examined and no 
flares of either the source or the background were detected.
The HRMA artifact\footnote{http://cxc.harvard.edu/ciao/caveats/psf\_artifact.html}, a feature identified in 
the central arcsecond of the \chandra\ PSF, 
was located, by using the tool {\it make\_psf\_asymmetry\_region}, in the SE direction (Fig. \ref{image} red region, top panel), 
opposite to the optical NW source position. 

Though accurate astrometry is already applied to \chandra\ data, we have run {\it wavedetect} 
to detect sources to be used for the fine relative alignment between the HRC-I and the COSMOS HST/ACS images. 
We selected the 4 sources at the smallest off-axis angles and with enough counts to properly determine their 
X-ray position. One of the 4 sources is a galaxy with extended optical morphology which has been discarded from the sample. 
Three sources were used to align the X-ray to the optical image. 
A small correction was applied in the y (0.06$^{\prime\prime}$) and x direction (-0.2$^{\prime\prime}$).
The final relative astrometry error is $\sim$0.01$^{\prime\prime}$.

\subsection{Data Analysis}
A single bright source is clearly visible in the unbinned data shown in Figure \ref{image} (top panel). The overlayed contours show the HST/ACS 
(green lines) smoothed image. 
The position of the SE optical source is co-incident with the peak of the X-ray emission, as also shown in Figure \ref{image} (bottom panel). 
No obvious X-ray source is present at the optical position of the NW source (marked with a cross in Figure \ref{image}). 

The number of net counts extracted in a 1$^{\prime\prime}$ radius circle is 379$\pm$20. Given that the spectral resolution of the HRC-I is very 
limited compared to ACIS and that the spectral shape of CID-42 has not changed in the past (C10), we assumed  
that the X-ray emission has a soft spectrum as observed in previous observations ($\Gamma=$2 and N$_{H,Gal}=2.7\times 10^{20}$ cm$^{-2}$)
to estimate the total flux of $\sim$9$\times$10$^{-14}$\cgs\ (L$_X=3 \times 10^{43}$\lum) in the 0.5-10 keV band. 

\subsubsection{2D Imaging Analysis}

In order to assess the presence of a second X-ray source and determine the flux ratio between the SE and NW optical sources, 
we performed a 2D imaging fitting. 
To model the PSF, we simulated a PSF image at the same off-axis angle of CID-42, 
by using Chart\footnote{http://cxc.harvard.edu/chart/index.html} and MARX\footnote{http://space.mit.edu/CXC/MARX/}. 
All fits were performed in {\it Sherpa} (Freeman et al. 2001) using modified Cash statistics and the Nelder \& Mead optimization method. 

\begin{figure}
\centering{\includegraphics[width=0.5\textwidth]{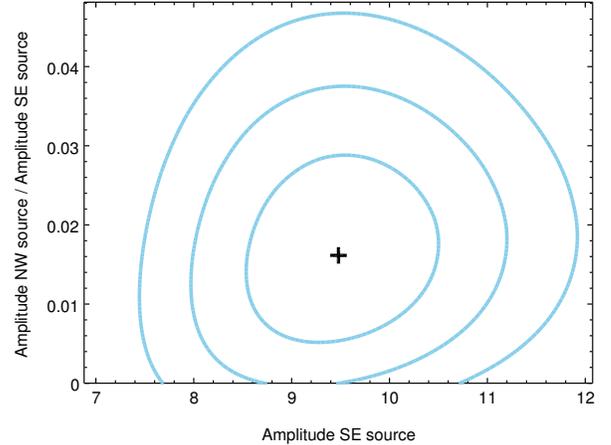}}
\caption{\small Confidence contours of fit statistic between the ratio between the amplitude of the two components and the amplitude of the 
brightest component. The 3 contours are drawn at 1, 2 and 3$\sigma$ confidence levels.}
\label{contours}
\end{figure}

We first measured the FWHM of the PSF by using a 2D Gaussian model, which well represents the PSF shape at the aimpoint. 
We then used a 3 component model to fit CID-42 X-ray emission: a constant background plus two 2D Gaussian components which FWHM matches the PSF. 
A family of fits was performed, using a grid of parameter starting points to
ensure the proper sampling of the multidimensional fit space.
We first left the centroid positions of the two Gaussian components free to vary. While the fitted centroid of the main source is consistent 
(within $<$0.2 pixel in the x direction and $\sim$0.7 pixel in the y direction) with the optical position of the SE source, the position of 
a second X-ray source is not constrained. 
We then fixed the position of the second Gaussian to be within few pixels from the optical position. We measured 
an upper-limit on the ratio between the amplitude of the two Gaussians of $<$4.5\% at 3$\sigma$ level 
(Fig. \ref{contours}), representing 
the ratio between the intensity of the two sources. 
By using the amplitude ratio measured with the 2D fitting, 
we find that the SE optical source is responsible for all  
of the X-ray emission, while the 3$\sigma$ upper limit on the contribution of the NW optical source is $\sim$4.5\% 
(17 out of 379 total counts), corresponding to an the X-ray total flux of F$_X < 3\times$10$^{-15}$ \cgs.

\subsubsection{Source Profile Analysis}

\begin{figure}
\fbox{\centering{\includegraphics[width=0.4\textwidth]{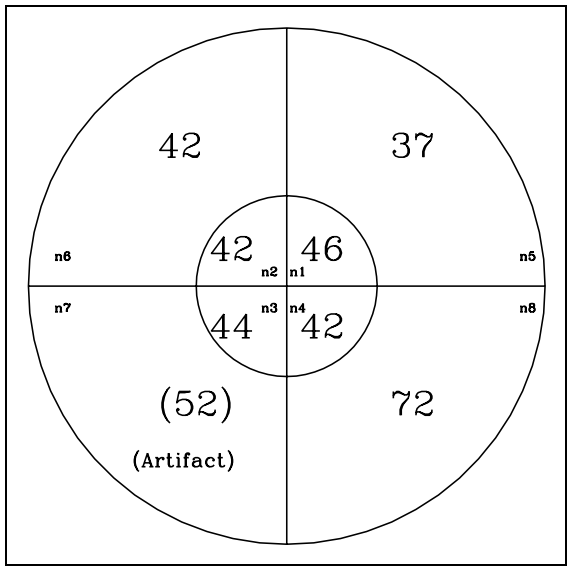}}}
\caption{\small The number of counts in each quadrant. The center correspond approximately to the SE source. 
Inner circle radius = 0.3$^{\prime\prime}$, outer circle radius = 1$^{\prime\prime}$.  }
\label{panda}
\end{figure}

Inspection of the X-ray image shows an elongation towards the SW direction (Figure \ref{image} middle panel). 
To quickly assess the significance of the asymmetry, we extracted the counts 
using the quadrant method (Elvis et al. 1983). 
We divided the region into a set of 8 inner and outer quadrants of radius 0.3$^{\prime\prime}$ and 1$^{\prime\prime}$, respectively. 
We moved the center around the position of the SE source slightly, until we found the location such that the most equal number of counts 
in each of the inner quadrants (n1 to n4) is measured. 
In Figure \ref{panda} the net counts in each quadrant are reported. 
In two of the outer regions (n5 and n6) we have a similar number of counts as expected from the symmetry of the PSF. The n7 quadrant is affected by 
the artifact region. 
Instead, there is a 5$\sigma$ excess of $\sim$32.5 counts in the region n8 with respect to region n5 and n6. 

\begin{figure*}
\includegraphics[width=0.5\textwidth]{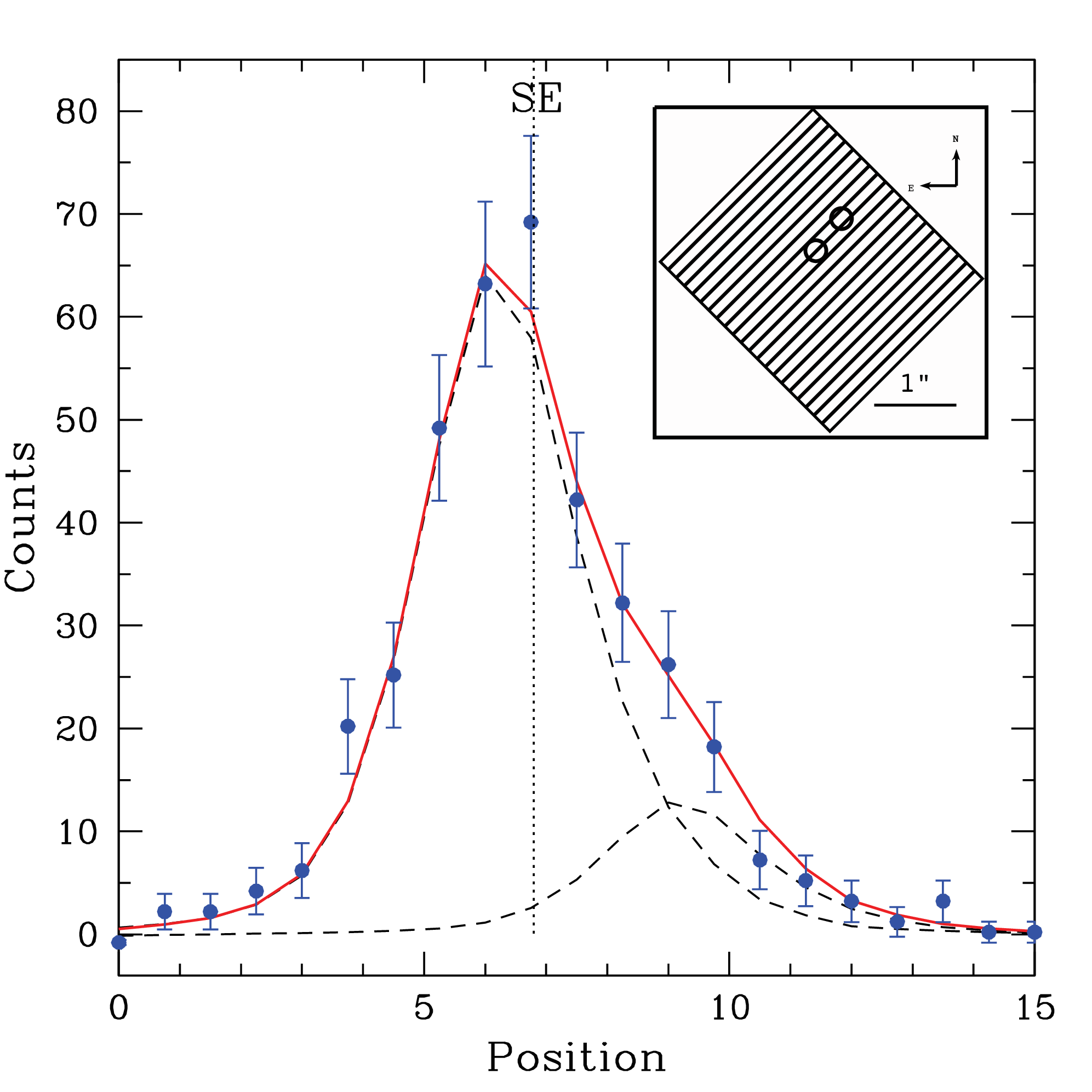}
\includegraphics[width=0.5\textwidth]{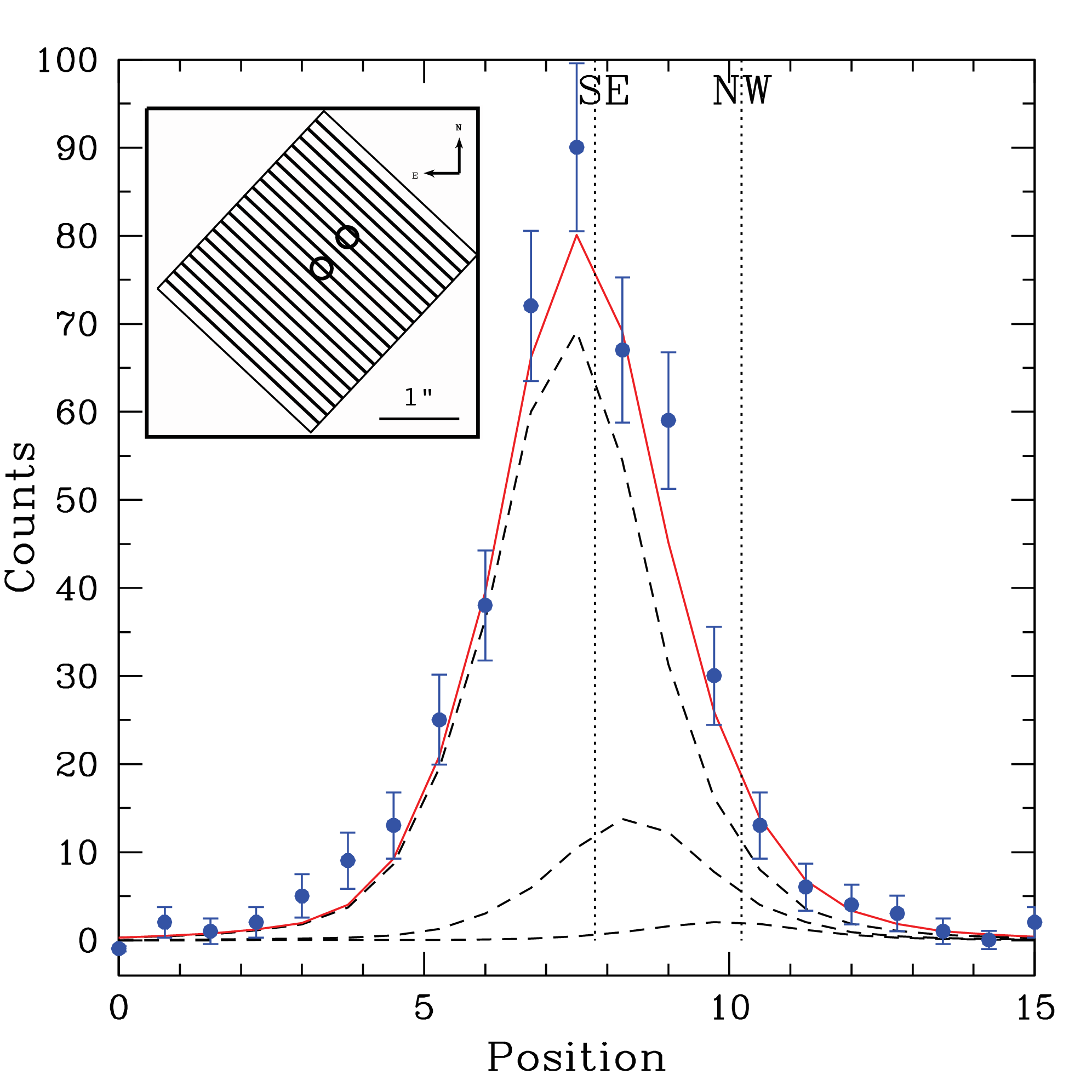}
\caption{\small Source profiles (blue filled circles) extracted in regions 
perpendicular to the line joining the two optical sources (left panel) and along the line joining the two optical sources (right panel). 
The dashed black lines represent the PSF profiles used to perform the fit. The solid red line represent the total fit. 
The position on the x-axis represent each 
extraction region of 0.16$^{\prime\prime} \times$2.5$^{\prime\prime}$ collapsed on 1D. The position of the SE and NW optical sources are marked.   
The extraction region grid (with the position of the NW and SE sources) is plotted in the sub-panel.}
\label{profile}
\end{figure*}

A more detailed fitting procedure was then employed. We extracted the counts using regions of 0.16$^{\prime\prime} \times$2.5$^{\prime\prime}$ (plotted in the panel enclosed in 
Figure \ref{profile}, left), to build the X-ray source profile (blue circles in Figure \ref{profile}, left panel) 
in a direction perpendicular to the line joining the two optical sources.
The profile shows a significant deviation from the PSF profile, extracted at the same position in the same direction (black dashed 
line in Figure \ref{profile}, left). 
By fitting the source profile with two PSF profiles we find that the amplitude of the SW X-ray feature 
contains $\sim$18\% of the amplitude of the main source. 

We also extracted another source profile along the line joining the optical sources (regions in the panel enclosed 
in Figure \ref{profile}, right). The source profile has 
the same shape of the PSF profile (in the same direction) but the intensity is higher than in the perpendicular direction.
When fitting the profile, a component to fit the SW X-ray feature is required 
with a peak intensity of $\sim$18\% of the main source, in agreement with the profile extracted in the perpendicular direction.
The SW feature is thus responsible for the larger number of counts at the peak of the radial profile along this direction. 
Instead, we can fit a component at the NW optical source, with intensity consistent with the finding of the 2D fitting 
($<$4\% of the main source), only by freezing its position. 

The SW X-ray feature source is offset $\sim$0.5$^{\prime\prime}$ ($\sim$2.5 kpc projected) from the center of the main point like source.
We performed extensive simulations (2000 runs) with MARX to determine whether the SW feature could be due to an irregular shape of the PSF. 
The probability that only a single point source is present is $<$0.005.

\section{Discussion}

The spatial resolution of HRC-I allows to confirm the presence in CID-42 of only one X-ray emitter and that the X-ray emission 
is centered on the SE source. A 3$\sigma$ upper limit is measured at the
position of the NW optical source.
In agreement with both recoil scenarios (GW and slingshot), the SE optical source has an X-ray emission consistent with being an 
unobscured active SMBH, as measured in the X-ray spectra in C10, and in agreement with 
the broad emission lines in the optical spectra (C10). 
 Obscuration from a torus cannot be present in the SE source in 
the recoil scenario of C10, because, when the SMBH recoils, only the most tightly
bound regions (disk and broad line regions) are carried with it, and the more distant 
regions (torus and narrow line regions) are left behind (Merritt et al. 2006; Loeb 2007;
Komossa \& Merritt 2008).
The broad emission lines and the unobscured X-ray spectrum of the SE source do 
not match well with the finding of Co09, who reported two narrow line (Type 2) AGN. 

The HRC flux measured for the SE  optical source is consistent, within the errors, with the flux 
observed in the 9 \xmm\ and \chandra\  observations ,taken over 5 years (C10 Figure 2, right). 
The upper limit estimated for the NW source is F$_X < 3\times$10$^{-15}$ \cgs, 
at least $\sim$50 and $\sim$15 times below the maximum (F$_X = 1.5\times$10$^{-13}$ \cgs in 2005) and the minimum 
(F$_X = 5\times$10$^{-14}$ \cgs in 2007) observed for CID-42.
The upper limit on the X-ray luminosity of the NW optical source is L$_X<10^{42}$\lum, which can be consistent either with emission from 
a normal or star-forming galaxy, or with a small fraction of scattered emission from a highly obscured AGN, or with a quiescent SMBH. 

Star formation rate (SFR) in the system could provide information on the nature of 
the X-ray emitter. We used the spectral energy distribution (SED), as published in C10, to compute the SFR.  
We normalized a standard Elvis et al. (1994) quasar 
template to the SE optical source magnitude measured in C10 and subtracted its contribution from the SED at all wavelength. We then performed a 
multi-component fitting using the $\chi^2$ minimization 
technique and several observed galaxies, quasar and hybrids (galaxy plus quasar) templates following the method of Ruiz et al. (2010) and 
Trichas et al. (2012, submitted). 
In the SED fitting we included also the SPIRE far-infrared data points from the Herschel Multi-tiered Extragalactic Survey 
(HerMES; Oliver et al. 2012 submitted) not available to C10. 
The best fit ($\chi^2/dof$=1.1) is produced by a young star-forming galaxy template (NGC 7714, Brandl et al. 2004, Ruiz et al. 2010) with a 
luminosity of L$_{8-1000 \mu m}$=6$\times$10$^{44}$ erg/s, SFR=25
M$_{\odot}$/yr, computed using the Kennicutt (1998) relation, and age of the
burst of 3-5 Myr. 
The statistical significance of the best fit is 9$\sigma$ above the second possible solution. The SFR derived and the infrared luminosity 
are consistent with the upper limit on the X-ray luminosity according to the 
relation between the 3 quantities derived for local galaxies (Ranalli et al. 2003). This fitting thus favors the presence of a young 
starforming galaxy in CID-42 in agreement with the measure of the X-ray upper limit for the NW optical source. 
However, the SFR rate measured could be produced not only by a young stellar 
cluster in the NW source, but could be the sum of the overall star formation in the system.

Since the total X-ray emission is dominated by the SE optical source, we cannot probe directly the presence of 
X-ray obscuration in the NW source from the X-ray spectral analysis.
An indirect probe of strong obscuration could be the presence of a strong (with an equivalent width of 1--2 keV) iron emission line (Matt et al. 2003). 
We indeed detected the FeK$\alpha$ line in CID-42 spectra (C10) with a rest frame equivalent width 
of EW=142$^{+143}_{-86}$ eV in the XMM-COSMOS spectrum (where the continuum was higher) and EW=$\sim$550$\pm$260 eV in the \chandra\ spectrum 
(where the continuum was lower), consistent with the EW typical of Seyfert 1 objects (Yaqoob \& Padmanabhan 2004). 
If we assume that the iron line observed were entirely produced by the NW optical source, which continuum emission is 
25 times fainter than the observed one, thus its EW should be a factor 25 higher, $~\sim$25 keV or more, an unphysical value.   

 A quiescent SMBH could still be hosted by the NW optical source. However,
observations in the local universe have shown that in gas-rich and or disturbed
systems (e.g., MW, NGC 1068, NGC 4151, Mrk 348), but also in early type galaxies
(Soria et al. 2006, Pellegrini 2010), active SMBHs are ubiquitous even if their activity could be sometime very low.

\subsection{The SW X-ray feature}

The SW X-ray feature has a flux of $\sim \times$10$^{-14}$\cgs\ (L$_X \sim 3\times10^{42}$ \lum) in the full HRC band.
The probability that this feature is a background source, at z$>$z$_{CID-42}$, within a 1.5'' radius circle from the center of CID-42 
is 8$\times$10$^{-5}$ (computed using the Gilli et al. 2007 number count predictions). 

From the X-ray and optical images and contours (Figure 1), a possible
association of the SW X-ray feature and
the minor SW optical tail in CID-42 is possible though this could just be due to a
projection effect.

Different physical scenarios can be used to explain the X-ray emission of the SW feature, each consistent with the recoil scenario: 
(1) Star formation could be the cause of the X-ray emission in the SW feature, with the same justification for the 
NW source (or, starburst-driven superwind activity). 
(2) The narrow [OIII] emission lines observed in the optical spectra of CID-42, at the same redshift of the 
host galaxy (see C10), could be produced by photoionization of the galaxy ISM by the ejected SMBH, on its way out
from the center. This would imply that the SMBH is close to a region of the galaxy disk with high ISM density 
or molecular clouds, to produce sufficiently high ionization.  The same
photoionized gas would also produce X-ray emission lines in the soft band
($\lesssim$1 keV), 
as already detected in some Seyfert galaxies (e.g., NGC~4151, Wang et al.
2011).
(3) A wind from the accretion disk (Elvis 2000) can produce bicones of gas 
moving at $\sim$500 km/s out from the nucleus. Internal shocks may heat enough of this gas to X-ray temperatures 
to be detectable. The luminosity would depend on the wind kinetic power. 
(4) The low L/L$_{Edd}\sim$0.04 estimated for the SMBH in CID-42 (C10) favors the presence of a jet, 
as in Galactic BH binaries where jets turn on at or below this Eddington ratio (Done et al. 2007). 
Though CID-42 is detected at 20 cm (f$_{20cm}$=138 $\pm$ 38 $\mu$Jy; VLA-COSMOS
survey Schinnerer et al. 2010), the low resolution (HPBW=1.5'') does not allow to 
associate the radio emission with the whole galaxy, the two optical sources or with the SW feature.
CID-42 is radio quiet ($R_L = log(f_{5GHz}/f_{B})$=-0.55, C10), but even radio quiet AGNs can have weak jets (Ulvestad \& Wilson 1984). 
Either a jet seen near our line of sight ($\lesssim$10 deg) as in e.g. 3C273 (Marshall et al. 2001), 
or a jet at a large angle (e.g., M87, Harris et al. 2009), of which we may be seeing only the brightest component, 
could produce the SW feature. 

\section{Future Observing Strategy}
 The new HRC X-ray data bring us to a more clear understanding of the nature of CID-42 SE optical source. 
Even if we now strongly prefer a star-forming nature for the NW optical source, 
the possibility of a truly Compton Thick AGN still remains open. The origin of
the newly discovered SW X-ray feature is still unclear. In the following we briefly
discuss observations that might shed even more light on CID-42. 

The major problem in reaching an unambiguous conclusion on the nature of the
optical emission line components at different velocities (i.e., broad emission lines and narrow
emission lines) is the uncertainty on their spatial location. This is technically challenging for
seeing-limited observations because of the small separation
(0.5$^{\prime\prime}$) between the two optical 
sources, thus either HST high-spatial resolution
spectroscopy or integral field spectroscopy with adaptive optics would be the
next step. 
As for example, we expect to see continuum blue emission and narrow emission
lines in the NW source if a cusp of young stars due to a starburst in the central 
gas reservoir is responsible for the upper limit measured in the X-ray band.

High resolution multi-frequency radio imaging is essential to locate and 
clarify the nature of the detected 20 cm radio emission in CID-42. 
Jansky-VLA observations in two different bands could
provide not only high angular resolution ($<$0.4$^{\prime\prime}$) to test for a jet in 
the SW X-ray feature, but also to separate the contribution of the nuclear 
and the stellar emission to the total radio flux.
Moreover, Very Long Baseline Array (VLBA) observations at 20 cm (PI: E.
Middelberg; final resolution of
15$\times$15 milli-arcsecond$^2$) are ongoing for all the VLA-COSMOS sources
including CID-42. 

New on-axis X-ray data with \chandra\ ACIS-S to get both the spectral and
spatial resolution (with the help of HRC data) could characterize the nature of the SW feature. Hard
X-ray emission is expected in the case of a jet, while soft band emission, due
to the presence of unresolved emission lines, is expected if the SW feature is
produced by photoionized gas. 

If a very obscured SMBH is hidden in the NW source, this could be detected by
NuSTAR (Harrison et al. 2010), which, in its 2-year prime mission, will devote 3.1 Ms to
mapping the C-COSMOS area in the 8-60 keV band.

\section{Summary}
We analyzed the high spatial  resolution HRC \chandra\ data obtained for CID-42, using both 2D spatial fitting and 1D radial profile fitting. 
Only 1 X-ray source is detected. 
The main X-ray emission is clearly produced by the SE optical source and the flux measured is consistent with 
previous observations.

A 3$\sigma$ upper limit on the X-ray emission of the NW optical source has been computed and its luminosity, together with the 
analysis of the multiwavelength SED, favor the presence of a star forming cluster in the NW source, instead of that of an obscured SMBH.
The unambiguous association of the X-ray emission with the SE optical source is consistent with the recoil scenarios proposed 
by C10 where only one active SMBH is expected. 
We note however, the presence of a very obscured active SMBH, as proposed by Co09, cannot be fully ruled out. 

Making use of all the observational properties measured for CID-42 (morphology, BH mass,
galaxy mass, luminosity, SFR, accretion rate), 
Blecha et al. (2012 submitted) will present a detailed modeling of CID-42 as both a GW recoiling SMBH and a
pre--merger, kpc--scale dual AGN. 

The discovery of a new X-ray feature, in a SW direction with respect to the main source, makes CID-42 even more intriguing. This feature 
could be connected with the star formation in the system, or emission from gas ionized by the SMBH or a jet.

 Future observations at the optical, X-ray and radio wavelengths will be proposed
to further investigate the peculiar properties of CID-42.

\begin{acknowledgements}
F.C. thanks A. Marinucci, A. Goulding, V. Kashyap and E. Schinnerer for useful discussion on the X-ray data. 
The authors thanks the referee for the useful suggestions which improved the
quality of the manuscript.
This work was supported by NASA Chandra grant
number GO7-8136A, the Blancheflor Boncompagni 
Ludovisi foundation and the Smithsonian Scholarly Studies. 
\end{acknowledgements}

\end{document}